\documentclass[letterpaper,10 pt,conference]{ieeeconf}  

\IEEEoverridecommandlockouts                              
\usepackage{graphicx}
\usepackage[table]{xcolor} 
\usepackage{graphics} 
\usepackage{color}

\title{\LARGE \bf
Putting Self-Supervised Token Embedding on the Tables
}

\author{Marc Szafraniec$^{1,2}$ Gautier Marti$^{1,3}$ Philippe Donnat$^{2,3}$
\thanks{Contact: marc.szafraniec@polytechnique.edu
        {\tt\small }}
\thanks{$^{1}$\'Ecole polytechnique, Route de Saclay, 91128 Palaiseau, France
        {\tt\small }}%
\thanks{$^{2}$OTCStreaming Ltd., Michelin House, SW3 6RD, London, UK
        {\tt\small }}%
\thanks{$^{3}$Hellebore Capital Ltd., Michelin House, SW3 6RD, London, UK
        {\tt\small }}%
}

\begin{document}

\maketitle
\thispagestyle{empty}
\pagestyle{empty}

\begin{abstract}

Information distribution by electronic messages is a privileged means of transmission for many businesses and individuals, often under the form of plain-text tables. As their number grows, it becomes necessary to use an algorithm to extract text and numbers instead of a human. Usual methods are focused on regular expressions or on a strict structure in the data, but are not efficient when we have many variations, fuzzy structure or implicit labels. In this paper we introduce SC2T, a totally self-supervised model for constructing vector representations of tokens in semi-structured messages by using characters and context levels that address these issues. It can then be used for an unsupervised labeling of tokens, or be the basis for a semi-supervised information extraction system.

\end{abstract}

\section{INTRODUCTION}

\begin{figure}[thpb]
  \centering
  \framebox{\includegraphics[width=0.35\textwidth]{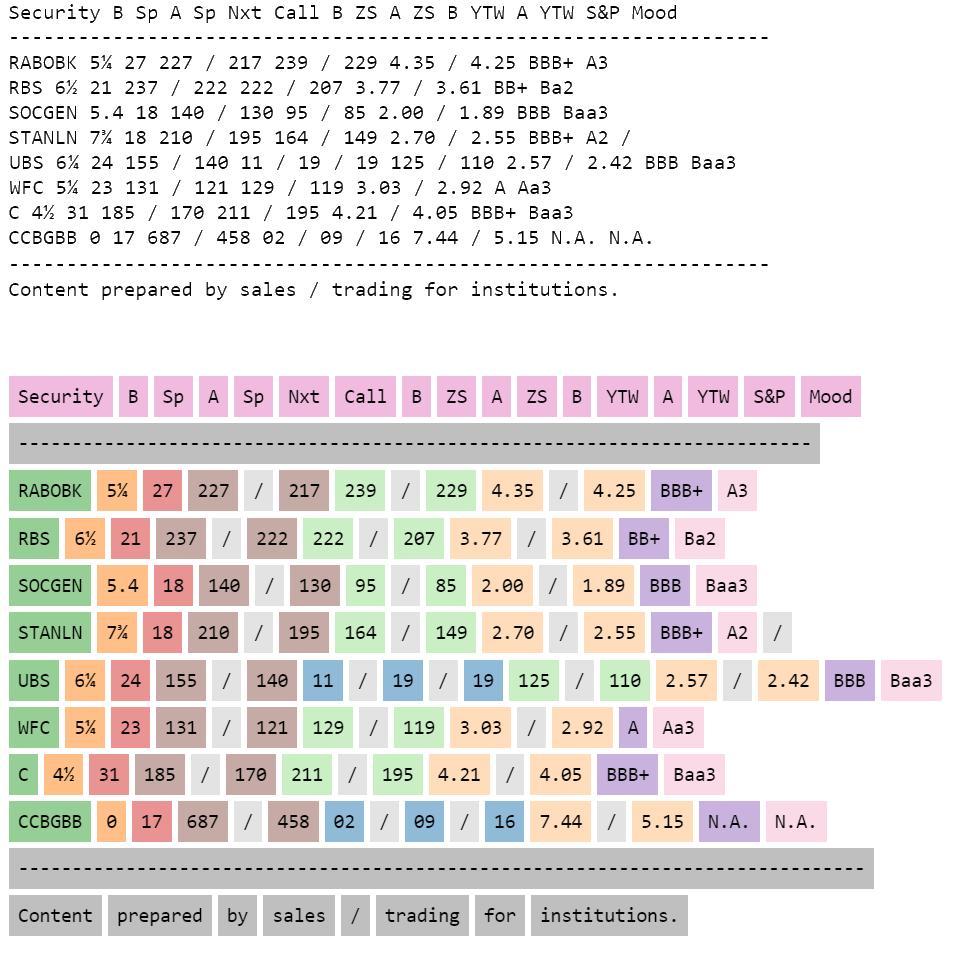}}
  \caption{An example of the type of ASCII table we want to extract, and the target extraction. The goal is to find what each token means, and each color corresponds to a type of token. We see that there are different line patterns, and this is only one type of message among thousands.}
  \label{asciiexample}
\end{figure}

Today most of business-related information is transmitted in an electronic form, such as emails. Therefore, converting these messages into an easily analyzable representation could open numerous business opportunities, as a lot of them are not used fully because of the difficulty to build bespoke parsing methods. In particular, a great number of these transmissions are semi-structured text, which doesn’t necessarily follows the classic english grammar. As seen in Fig.~\ref{asciiexample}, they can be under the form of tables containing diverse elements, words and numbers, afterwards referred to as \textit{tokens}. 

These tables are often implicitly defined, which means that there are no special tags between what is or not part of the table, or even between cells. In these cases, the structure is coming from space or tabs alignment and from the relative order of the tokens. The data often are unlabeled, which means that the content must be read with domain-based knowledge. Thus, automatic extraction of structured information is a major challenge because token candidates come in a variety of forms within a fuzzy context. A high level of supervision is hard to obtain as manual labeling requires time that is hardly affordable when receiving thousands of such emails a day, and even more so as databases can become irrelevant over time. That is why training a generalizable model to extract these data should not rely on labeled inputs, but rather on the content itself - a paradigm called \textit{self-supervised learning}. Many approaches already exist in Natural Language Processing, such as Part-of-Speech (POS) tagging or Named Entity Recognition (NER), but they do not take advantage of the semi-structured data framework. On the contrary, there exists some information extraction algorithms applied to tables, but they necessitate a great amount of manually defined rules and exceptions. Our model aims to reconcile both approaches for an efficient and totally self-supervised take on information extraction in the particular context of semi-structured data.

In this paper, we present a neural architecture for token embedding in plain-text tables, which provides a useful lower-dimensional representation for tasks such as unsupervised, or semi-supervised clustering. Intuitively, tokens with a similar meaning should be close in the feature space to ease any further information extraction. Our model aims to combine the better of the context and the character composition of each token, and that is why the neural architecture is designed to learn both context and character-level representations simultaneously. Finally, we can take advantage of the distances between tokens in the feature space to create proper tables from fuzzy input data.

\section{RELATED WORK}

\subsection{Information Extraction on Semi-Structured Data}
\label{rw:iessd}

The field of Information Extraction on Semi-Structured Data has been particularly active in the 1990's and the early 2000's, developed in settings such as the Message Understanding Conferences (MUCs) and, more recently, in the ICDAR 2013 Table Competition \cite{gobel2013icdar}. A very complete survey of information extraction in tables can be found in \cite{turmo2006adaptive} and in \cite{embley2006table}. The main goal of systems such as \cite{pinto2003table}, \cite{mccallum2000maximum} or TINTIN \cite{pyreddy1997tintin}, is to detect tables in messages, or to label lines such as captions using the density of blank spaces, Conditional Random Fields or Hidden Markov Models respectively. This also has been done more recently in an unsupervised manner by \cite{cortez2013unsupervised} and \cite{yeh2013unsupervised}. Obviously the main goal is to extract the content of these tables, which is done by \cite{ciravegna2001adaptive,viola2005learning,tengli2004learning,soderland1997learning}, with DEByE \cite{laender2002debye}, DIPRE \cite{Brin1999} or WHISK \cite{soderland1999learning} by learning patterns to match to the data systematically using manually defined rules and trying to generalize them as much as possible. A very thorough panorama of this class of algorithms is presented in \cite{chang2006survey}. More recently, \cite{kasar2015table} proposes a graph structure in tables to match predefined patterns. Unfortunately, these methods are not flexible enough to be used in the case of a great number of patterns in the data, and need user supervision or gazetteers to work properly, which are not always available. The idea of our model can certainly be related the most with \cite{agichtein2004mining} or \cite{mintz2009distant}, but we add in new Natural Language Processing tools and neural networks -- among other differences.


\subsection{Natural Language Processing}

In recent years, neural networks have replaced handcrafted features in Natural Language Processing, with excellent results -- a recent survey of the topic can be found in \cite{lopez2017deep}. The seminal paper of Collobert et al. \cite{collobert2011a} presents a first idea of token embeddings, or word features vectors, based on lookup tables in a fixed vocabulary and using neural networks. It also brings a general solution to problems such as Part of Speech (POS), Chunking and Named Entity Recognition (NER). The work on word features vectors continued with the classic Word2Vec paper \cite{mikolov2013distributed} which is now one of the references on the topic, introducing the skip-gram model for text. There, the method used to train the network is trying to predict the next words in a sentence based on surrounding ones. However, a problem of these approaches are that they rely on a dictionary of words, and that ``out-of-vocabulary'' words such as orthographic errors get a generic representation. In problems such as information extraction, that is a major issue because the content consists mostly in names that are not classic words, and can evolve in time. Besides, closely related words such as ``even'' and ``uneven'' should be close in the feature space, which is not guaranteed by these methods. That is why recently the focus has shifted on a study directly on the characters, that mostly solve these questions. Examples can be found in \cite{ling2015finding} and \cite{lample2016neural} with LSTMs, or in \cite{kim2016character}, \cite{chiu2015named} and \cite{santos2014learning} with Convolutional Networks. Further developments presented in \cite{li2015hierarchical} and \cite{le2014distributed} aim to learn vector representations of sentences or documents instead of limiting the models to the words only. This is done with the same methods used to get words representations, only with whole rows or paragraphs as the input. These are our main inspirations, but all these algorithms have been created to deal with natural and not semi-structured text, so they do not take advantage of the bi-dimensional structure of the data. An effort worth noting is \cite{graves2009offline} with the introduction of Multidimensional Recurrent Neural Networks in the Optical Character Recognition (OCR) field, but the idea has not been developed further.









\section{THE SC2T EMBEDDING}

We will now present the SC2T (Self-Supervised Character and Context-levels on Tables) embedding. As in \cite{lample2016neural}, two important ideas guide our neural network architecture: to correctly represent a token, we need to take into account its composition (a number, a word?) as well as its context (the surrounding tokens). As we deal with tokens that mostly are not words in the classic sense of the term, but abbreviations, numbers, unique identifiers... and that we have no dictionary, we can't use word-level features similar to what was done in \cite{mikolov2013distributed}. That's why we will use character-level representations, in the same fashion that \cite{lample2016neural}, \cite{kim2016character}, \cite{ling2015finding} or \cite{santos2014learning}. We do not use external dictionary or gazetteers, which allows our program to be relevant on any semi-structured text. Note that given raw text as input, the first stage is the tokenization of the data. A discussion on that topic is complex and beyond the scope of this paper, as special rules have to be applied depending on the data and pertinent segmentation.

\subsection{The Architecture}
\label{sc2t:arch}

\begin{figure}[thpb]
  \centering
  \framebox{\parbox{3in}{\includegraphics[width=0.4\textwidth]{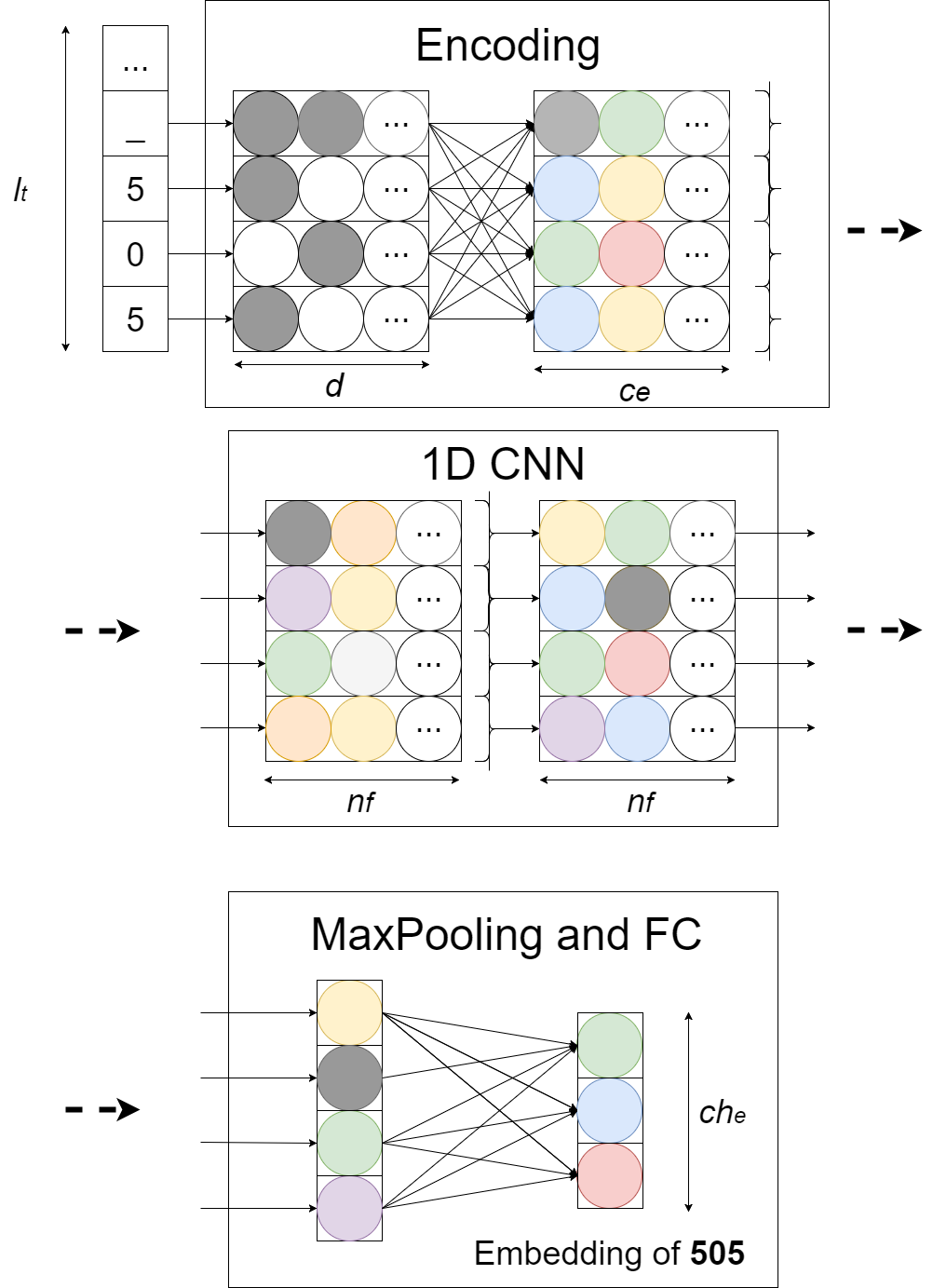}}}
  \caption{Illustration of the character-level neural network}
  \label{fig:chnn}
\end{figure}

\begin{figure}[thpb]
  \centering
  \framebox{\parbox{3in}{\includegraphics[width=0.44\textwidth]{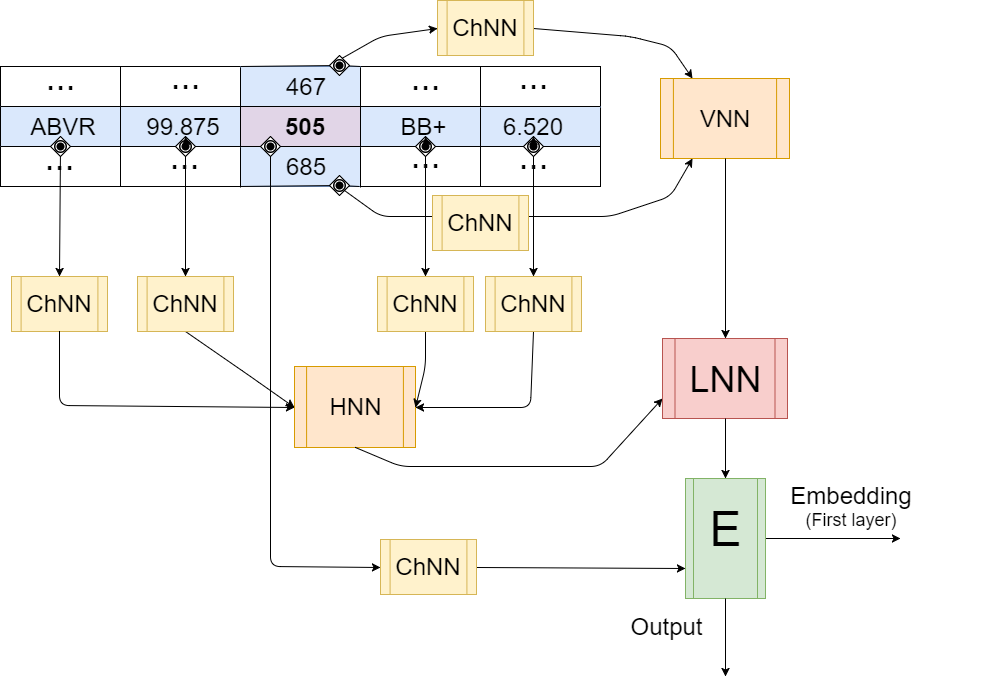}}}
  \caption{General schema of the model architecture and the generation of an embedding (\textbf{E}) for the middle token (\textbf{505}) -- $h_w = 5, v_w = 3$}
  \label{gendiagram}
\end{figure}

Our architecture is created to learn a character- and context-sensitive embedding of tokens. To build this distributed representation we train our network on a proxy task, which is to reconstruct tokens using only the surrounding ones - an idea recalling auto-encoders. By \textit{surrounding}, we mean that are contained in a horizontal window of size $h_w$ and a vertical window of size $v_w$ around it, padding with zeros if necessary. This method resembles what is done in \cite{collobert2011a} or \cite{le2014distributed} for example, but takes advantage of the 2D structure of the data. Selecting tokens which are horizontally adjacent is trivial contrary to vertical ones. Papers such as \cite{trecs1} and \cite{trecs2} give good insights on how to define that efficiently. However, for simplicity reasons, we take the tokens of the surrounding lines which rightmost character is closest to the rightmost character of our target token. Each of these surrounding tokens is first transformed in a one-hot encoding on the characters of dimensionality $d$, padded left with blank spaces to achieve the length $l_t$ for all tokens. Then, they all pass in the same character-level convolutional network (ChNN), which structure is inspired by \cite{santos2014learning}. It is composed of a one-hot-encoding then fully connected (FC) layer, then of two one-dimensional CNNs with $n_f = 64$ filters of size $3$ with a max-pooling. Finally, a fully connected layer is added to bring the embedding to the desired size. ReLU activations, batch normalization and $25\%$ dropout are also placed between each layer. A diagram of this network can be found in Fig.~\ref{fig:chnn}.
The resulting embeddings are then concatenated and fed into the horizontal (HNN) and vertical (VNN) context networks, that have the same structure as the character-level network excepted the input size and that the max-pooling and FC layer is replaced by a simple Flatten layer. They are kept separate from each other because they are not aimed to learn the same relationships in the data. Then their outputs are merged and passed through two fully connected layers (LNN), the last of them of size $s_e$. Thus, we have two useful representations for a given token: the output from the LNN network (of size $s_e$), plus the output taken directly from the character CNN on the token itself (of size $ch_e$). We then concatenate and feed them to the last part of the network, E, which consists of two fully connected layers and whose final output is compared to the one-hot-encoding of the original token. The concatenation is followed by a dropout layer to prevent the network to only use the input token. A value of $0.5$ yields the best results in our experience, which confirms the idea presented in \cite{lample2016neural}. Our model allows a simultaneous training of all the components in the network using backpropagation. Finally, our context- and character-sensitive embedding is obtained by taking the output of the first FC layer in the E network, which has size $ch_e + s_e$, and we will see in the next part that it is indeed a useful distributed representation of tokens. A diagram of our whole network can be found in Fig.~\ref{gendiagram}.

We use CNNs in all the stages of our network instead of LSTMs or other layers for two reasons: first, in the case of tables, the sequential aspect is often negligible. Besides, we implemented the same program with bidirectional LSTMs and it did not yield better results, while slowing down the whole process. This is a problem because speed of execution is an important factor in industrial applications treating tens of thousands of messages each day, each containing hundreds or thousands of tokens.

\subsection{Alternative Model}
\label{sc2t:am}

An alternative to the previous model can be considered. Indeed, instead of letting the E network merge the character and context embeddings, we could just concatenate them, applying a constant importance coefficient $K$ that has to be defined depending on the data. Indeed, if the different categories in the data are from different types (e.g., textual names and numbers), the character content has to be privileged, unlike the case of more context dependent tokens (e.g., numbers in a certain order). Usually, if the structure of the data is disrupted, we will need to rely more on characters. $K$ will increase the weight of one part or another, given that clustering algorithms put more importance on greater values in the data. Obviously, this coefficient $K$ necessitates an intervention of the user, and a knowledge of the data. Thus, it is not applicable in general but can be very efficient in particular cases, as we will see in section IV.

\subsection{Tokens and Lines Clustering}
\label{sc2t:lcl}

Once we obtain our token embeddings, a simple clustering algorithm such as \textit{k-means++} \cite{arthur2007k} can be used to compute a clustering of the tokens. Obtaining coherent groups of tokens can lead to many developments. It can be used for manual labeling and bootstrapping quickly a labeled dataset for supervised learning, but it can also be the basis of an efficient semi-supervised algorithm.

We also need to cluster lines in the data: indeed, a message is often composed of one or multiples headers, the data itself, as well as disclaimers and signatures, and more generally blocks of natural language in the document. Once again, their repartition or presence is not guaranteed, so an adaptable clustering is necessary. To obtain an embedding of the lines, we simply compute a max-pooling of the embeddings of its tokens. We used this method for separating headers, disclaimers and table content by \textit{3-means} clustering on our data.

\section{EMPIRICAL RESULTS}

To assess the efficiency of our embeddings, we use them to label tokens in the Online Retail Data Set from UCI\footnote{http://archive.ics.uci.edu/ml/datasets/online+retail} via \textit{k-means++} clustering. We chose it because this is a varied public dataset that fits the kind of problem we are dealing with. Unfortunately, the relevant Information Extraction papers we found (sec. \ref{rw:iessd}) used either custom datasets, or datasets that are not online anymore.

\subsection{The Dataset}

\begin{figure}[thpb]
  \centering
  \framebox{\parbox{3in}{\includegraphics[width=0.40\textwidth]{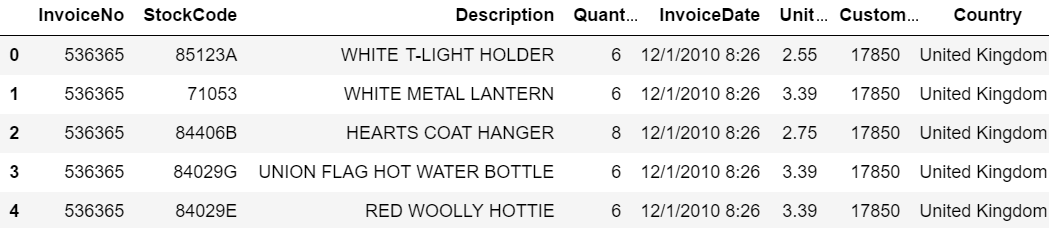}}}
  \caption{The first few lines of the dataset}
  \label{firstfew}
\end{figure}

The Online Retail Data Set consists of a clean list of $25873$ invoices, totaling $541909$ rows and $8$ columns. \textit{InvoiceNo}, \textit{CustomerID} and \textit{StockCode} are mostly 5 or 6-digit integers with occasional letters. \textit{Quantity} is mostly 1 to 3-digit integers, a part of them being negative, and \textit{UnitPrice} is composed of 1 to 6 digits floating values. \textit{InvoiceDate} are dates all in the same format, \textit{Country} contains strings representing 38 countries and \textit{Description} is 4224 strings representing names of products. We reconstruct text mails from this data, by separating each token with a blank space and stacking the lines for a given invoice, grouped by \textit{InvoiceNo}. We will use the column label as ground truth for the tokens in the dataset. For simplicity reasons we add underscores between words in \textit{Country} and \textit{Description} to ease the tokenization. Another slight modification has to be done: $25\%$ of the \textit{CustomerId} values are missing, and we replace them by '00000'. A sample can be found in Fig.~\ref{firstfew}.

\subsection{Labeling of tokens using the SC2T Embedding}

We will now create an embedding of the tokens, and use it in a \textit{k-means++} clustering. We will use the \textit{homogeneity score} $h$ as metrics, which measures if all the data points that are members of a given cluster are given the same label. It can be written

\[h = \frac{1}{k}\sum_{c = 1}^{k}\frac{\#~C_c \cap L_c}{\#~C_c}\]

where $C_c$ is the ensemble of data points in cluster $c$ and $L_c$ is the ensemble of data points that have the label which is most present in cluster $c$. It represents the accuracy of a semi-supervised clustering where the user simply gives a label to each cluster, corresponding to the majority of its elements. Obviously, $h\to 1$ when $k$ tends to the number of data points. However, we will not restrain ourselves to taking $k = 8$, the exact number of labels, as varied data can have the same ground truth labels in a real setting. For example, $12/24/2017$, $2017$ or $Dec-24$ could be all labeled as dates, but might be difficult to group into one cluster. That is why we do not consider the \textit{completeness score}, which measures if all the data points of a given class are elements of the same cluster, as relevant in our case. So, a good measure of the quality of our clustering is the score reached for a certain number of clusters, e.g. $20$ or $100$, which will represent the number of points that the user should label to obtain such accuracy. Note that as \textit{k-means} yields stochastic results, the results given here are a mean of $100$ independent runs.

\begin{figure}[thpb]
  \centering
  \framebox{\parbox{3in}{\includegraphics[width=0.40\textwidth]{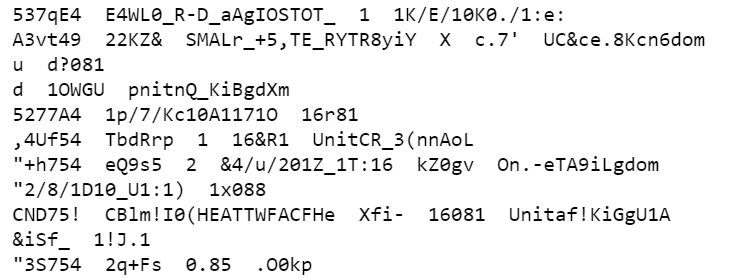}}}
  \caption{A few input lines in our most difficult setting (Del. 50\%, CR 50\%)}
  \label{difficult}
\end{figure}

At first, we have a simple problem: all the lines follow the same pattern, so a simple extraction rule can perfectly extract data. This is a good baseline for our program as it should retrieve all the information. Our experiment consists of creating homogeneous clusters according to the labels of the tokens after randomly deleting a portion of them (\textit{Del.}) and/or replacing randomly a part of the characters (\textit{CR}) - heavy modifications that are not unlike those found in real-life settings. An example of disrupted data can be found in Fig.~\ref{difficult}.
Note that we only used a subset of $1000$ invoices, $24K$ lines or approximately $190K$ tokens, which yielded slightly worse results compared to the tests we made on the whole dataset. It is logical that the more the context is disrupted, the more we will rely on the characters part. We will present the results in two settings: one with the model presented in \ref{sc2t:arch} \textit{(NoK)}, the other one with the parameter $K$ presented in \ref{sc2t:am} \textit{(K)}. \textit{Best Char \%} is the proportion of the norm of the character part of the embedding compared to the norm of the whole embedding, which is controlled by variations of $K$. Results of homogeneity depending on the number of clusters can be found in Table I ($nc$ being the number of clusters), and our parameters in Table II. We chose the horizontal window such as it takes into account the whole line, but that could be unadapted in the case of very large tables. 

\rowcolors{2}{gray!15}{white}
\begin{table}[!h]
\caption{Homogeneity of the Clustering\vspace*{-1em}}
\label{accuracy_clustering}
\begin{center}
\begin{tabular}{lccccc}
\hline
$nc$ & & 8 & 20 & 100 & Best Char \%\\
\hline
Full Data & NoK & 84.1 & 94.7 & 99.3 & --\\
& K & 100 & 100 & 100 & 0\\
Deletion 5\% & NoK & 86.9 & 95.2 & 98.0 & --\\
& K & 97.8 & 97.2 & 99.3 & 80\\
Deletion 10\% & NoK & 87.5 & 95.9 & 98.4 & --\\
& K & 93.3 & 96.5 & 98.7 & 90\\
Deletion 50\% & NoK & 78.0 & 89.9 & 96.7 & --\\
& K & 92.4 & 95.9 & 97.7 & 100\\
Char. Repl. 5\% & NoK & 99.3 & 99.7 & 100 & --\\
& K & 100 & 100 & 100 & 0\\
Char. Repl. 50\% & NoK & 60.0 & 73.1 & 93.0 & --\\
& K & 99.9 & 99.9 & 100 & 0\\
Del. 10\% + CR 10\% & NoK & 73.2 & 92.2 & 97.3 & --\\
& K & 88.9 & 93.2 & 97.1 & 80\\
Del. 10\% + CR 50\% & NoK & 62.3 & 76.1 & 94.2 & --\\
& K & 71.7 & 80.9 & 90.3 & 20 \\
Del. 50\% + CR 10\% & NoK & 76.5 & 88.3 & 94.7 & --\\
& K & 89.4 & 91.0 & 94.7 & 90\\
Del. 50\% + CR 50\% & NoK & 70.2 & 81.6 & 88.4 & --\\
& K & 64.3 & 69.2 & 74.6 & 80\\
\hline
\end{tabular}
\end{center}
\end{table}

\begin{table}[h]
\caption{Values of the parameters\vspace*{-1em}}
\label{value_parameters}
\begin{center}
\begin{tabular}{rcc}
\hline
\rowcolor{white}
PARAMETER & NAME & VALUE\\
\hline
Character Dictionary Dim. & $d$ & $78$\\
Context Embedding Dim. & $s_e$ & $50$\\
Character-Level Embedding Dim. & $ch_e$ & $50$\\
Max. Length of Tokens & $l_t$ & $20$\\
Horizontal Window & $h_w$ & $20$\\
Vertical Window & $v_w$ & $5$\\
\hline
\end{tabular}
\end{center}
\end{table}

Obviously, the more disrupted the data, the less accurate our model. First, we can see that the model with $K$ is better than without in most cases, but remember that the value of $K$ has been cross-validated to obtain the best possible result. This is not realistic in general, but can still be very useful when we have prior knowledge about the data. For example, we observe that without deletions and even with character replacements, the context alone brings 100\% accuracy, reflecting that the position entirely determines the label. When we randomly replace characters we cannot rely as much on them, and numbers show that our model is more robust to a deletion of tokens than it is to character replacement, probably because in our dataset tokens with the same label are often similar in composition. It is also interesting to notice that our supervision-free \textit{NoK} model, even if slightly disadvantaged in simple cases, yields its best results when the data is more disrupted. This is good news, as it is in these cases that we have the least amount of prior knowledge, besides being certainly the most realistic settings and the ones that need new models most.

Without surprise, we noticed that it is often \textit{CustomerID}, \textit{InvoiceNo} and to a lesser extent \textit{StockCode} that are mislabeled, due to their same composition. Even in our most difficult case, 50\% deletion and 50\% character replacement, we obtain decent results in our unsupervised setting. Overall, with as few as $100$ token labels out of $190K$ we could get a high clustering accuracy on most of our contexts. The size of the embedding also had to be chosen carefully, because it has to encode enough information while avoiding the curse of dimensionality. Finally, note that the network gets less training data when increasing the percentage of deletions, and that we retrained it from scratch in each setting.

\subsection{An Application to Table Alignment}

\begin{figure}[thpb]
  \centering
  \framebox{\parbox{3in}{\includegraphics[width=0.40\textwidth]{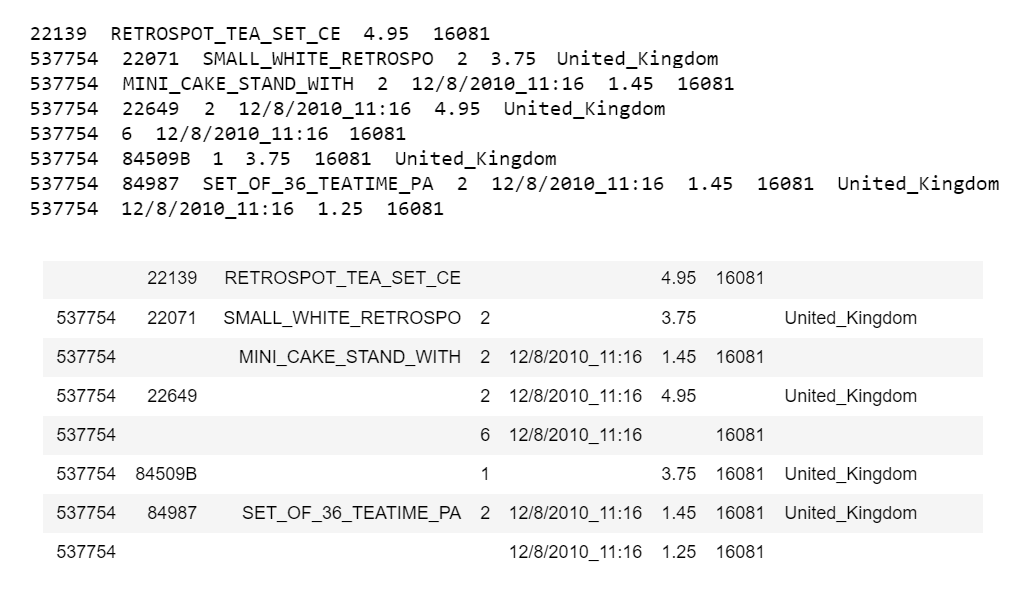}}}
  \caption{Input test message and aligned table after $30\%$ deletion. The penultimate line is the reference here, as it is the most complete one.}
  \label{alignment}
\end{figure}

Often, tables are not correctly aligned when data is missing, which creates an erroneous display. To correct this problem, we can define a reference line, that is the longest line that belongs to the table part according to the lines clustering. This line will define the number of columns in our resulting table. Then, for every other line, we try to match each token with a token from the reference line that is on its right, i.e. the token which is closest in the embedding space while allowing the order to be kept. We suppose here that the order is always preserved because in a given table permutations are very unlikely. We then obtain correctly aligned tables, as seen in Fig.~\ref{alignment}, which can be very useful for an easier labeling of the tokens. This can be used even if there are different types of lines containing different information, theses lines being separated beforehand by clustering as presented above in \ref{sc2t:lcl}. We then take different rows as references.

\section{CONCLUSIONS}

In this paper we present a new Neural Language model that jointly uses the character composition of tokens and their surrounding context in the particular framework of semi-structured text data, for the purpose of generating a distributed representation. We have seen that the embeddings have linearized the space quite well such that a $k$-means will gather similar tokens, or by max-pooling them, similar lines, and that it could be applied to table realignment. The approach presented here can already allow an information extraction system to function, but it could be even more beneficial to add semi-supervised learning algorithms, as described in \cite{yu2016incremental} or \cite{calandriello2016incremental}. Another solution would be to bootstrap large annotated databases for performing supervised learning. We introduce several hyper-parameters to be tuned, mainly the sizes of our embeddings. We want our model to stay as general and unsupervised as possible, and we argue that tuning them manually is the better solution as existing unsupervised measures of the quality of a clustering (Silhouette Coefficient \cite{rousseeuw1987silhouettes}, Calinski-Harabaz Index  \cite{calinski1974dendrite}) can be misleading for our particular task. Indeed they can favor less clusters that are not homogeneous in terms of labels instead of more cluster that are, which is against our goal. Finally, the fact that we do not have relevant standards for this particular task is problematic. However, our dataset is openly available on the Internet (link above), and can be a simple but representative benchmark for papers to come.




\vspace*{1em}

\section*{ACKNOWLEDGMENT}

We would like to thank Clement Laisn\'e (Hellebore Technologies) for having developed 
convenient tools that greatly helped us in our research, 
as well as all our colleagues for their support.
We also thank Caio Filippo Corro for discussions about this paper.

\vspace*{2em}

\bibliographystyle{IEEEtran}
\bibliography{references}

\end{document}